# Balanced multi-shot EPI for accelerated Cartesian MRF: An alternative to spiral MRF

Arnold Julian Vinoj Benjamin[1,2], Pedro A. Gómez[3,4], Mohammad Golbabaee[1], Tim Sprenger[4], Marion I. Menzel[4] Michael Davies[1] and Ian Marshall[2]

[1]School of Engineering, Institute for Digital Communications, University of Edinburgh, [2]Centre for Clinical Brain Sciences, University of Edinburgh, [3]Computer Science, Technische Universität München, Munich, Germany, [4]GE Global Research, Munich, Germany

**Synopsis:**

This study shows the practical implementation of an accelerated Cartesian Magnetic Resonance Fingerprinting (MRF) scheme using a multi-shot Echo Planar Imaging (EPI) readout. Its performance is compared with conventional spiral MRF and the fast convergence of accelerated iterative reconstructions for this method is shown.

**Purpose:**

The main purpose of this study is to show that a highly accelerated Cartesian MRF scheme using a multi-shot EPI readout (i.e. multi-shot EPI-MRF) can produce good quality multi-parametric maps such as T1, T2 and proton density (PD) in a sufficiently short scan duration that is similar to conventional MRF [1]. This multi-shot approach allows considerable subsampling while traversing the entire k-space trajectory, can yield better SNR, reduced blurring, less distortion and can also be used to collect higher resolution data compared to existing single-shot EPI-MRF implementations [2, 3]. The generated parametric maps are compared to an accelerated spiral MRF implementation with the same acquisition parameters to evaluate the performance of this method [4]. Additionally, an iterative reconstruction algorithm is applied to improve the accuracy of parametric map estimations and the fast convergence of EPI-MRF is also demonstrated [5].

**Methods:**

The scanning was performed on a GE HDx 3T MR750w scanner with a 12 channel receive only head RF coil (GE Medical Systems, Milwaukee, WI). 16-shot EPI-MRF datasets and spiral datasets with 89 interleaves and golden angle rotations were acquired from a tube phantom (Diagnostic Sonar, Livingston, UK) consisting of tubes with different T1 and T2 values and a healthy volunteer using the inversion recovery (IR) prepared ($T_{inv}$ = 18 ms) Quantitative Transient-state Imaging (QTI) sequence, using a linear ramp flip angle (FA) variation from 1° to 70° for 500 frames [4]. The $G_x$ and $G_y$ gradient of the multi-shot EPI trajectory were balanced (Fig. 1) to ensure that the residual magnetization remained constant for every shot and the crusher gradient ($G_z$) was applied to introduce gradient spoiling. The TR was set to 16 ms for all the datasets to enable comparison between multi-shot EPI-MRF and spiral MRF. The acquisition time for a single slice was 9 s. Both acquisitions had 22:5 x 22:5 cm Field of View (FOV), 128 x 128 matrix size, 1.3 mm in-plane resolution and 5 mm slice thickness. The reconstruction was performed by the CS based dictionary matching method with iterative reconstructions to generate quantitative T1, T2 and PD maps [5, 6]. The dictionaries were calculated using the Extended Phase Graph (EPG) model [7].

**Results:**

Fig. 2 shows the highly subsampled aliased images of the tube phantom and healthy volunteer at different frames/repetitions indexes t along with the CS-based reconstruction that removed the aliasing and provided a better visualization of the signal temporal dynamics. Fig. 3 shows the T1 and T2 sensitivity of the sequence for discriminating dictionary elements using a linear ramp FA variation from 1° to 70° for 500 frames. Fig. 4 shows the generated T1, T2 and PD maps for the tube phantom and healthy volunteer for both the 16 shot EPI-MRF and spiral MRF acquisitions. Fig. 5 shows the qualitatively improved parametric estimations of T1, T2 and PD maps for the tube phantom and healthy volunteer after the application of an iterative projection algorithm.

**Discussion and Conclusion:**

Fig. 3a shows that the T1 sensitivity is high throughout the acquisition, is enhanced by the initial inversion pulse and occurs mostly at lower flip angles whereas Fig. 3b shows that the T2 sensitivity occurs mostly at higher flip angles (> 25°). The T1 and T2 values of EPI-MRF and spiral MRF in Fig. 4 and Fig. 5 for grey matter (GM) and white matter (WM) are very similar to each other, in close agreement to those reported in literature [8]. However, there is an underestimation of cerebrospinal fluid (CSF) T2 values in both EPI-MRF and spiral MRF and slight aliasing artefacts also appear in the T2 and PD maps of EPI-MRF. This is because the encoding scheme used in the acquisition is comparatively less sensitive to T2 variations than T1 variations. The use of an optimized FA train instead of a linear ramp may yield more accurate T2 values and may potentially suppress the ghosting artefacts in the T2 and PD maps.

Fig. 5 shows that the accuracy of the parametric maps was improved by the use of iterative reconstruction algorithms. The iterative reconstructions of multi-shot EPI-MRF converges very quickly (35 s) compared to spiral MRF (~ 4 minutes) and could therefore result in a very fast implementation on the scanner. This could be further improved by the use of an adaptive iterative algorithm [9]. The convergence of spiral acquisition is slow (which means more iterations) because spiral sampling is ill-posed [10]. Moreover, each iteration is more expensive because spiral sampling uses costlier non-uniform fast Fourier Transform (NUFFT) compared to FFT in EPI. In addition, higher resolution data can be acquired using a multi-shot EPI acquisition because it does not suffer from blurring artefacts that become more pronounced in spiral acquisitions at longer readout durations [11].

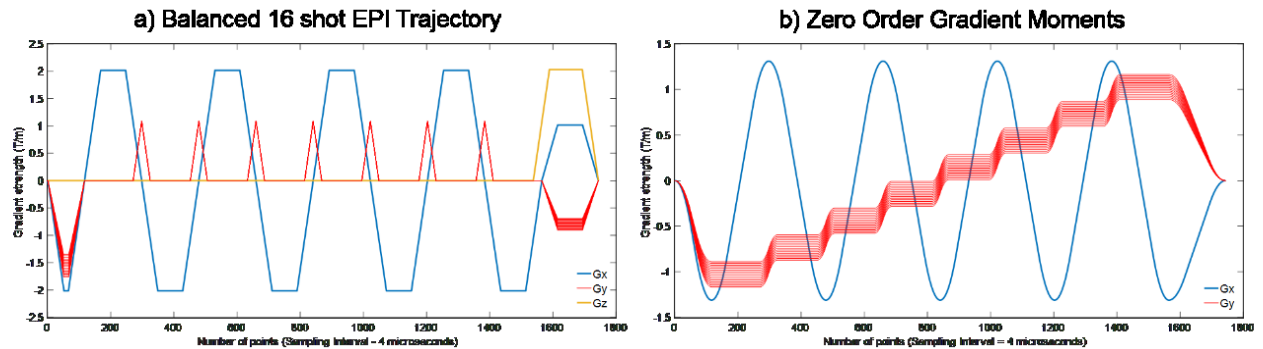

Figure 1: The balanced 16 shot EPI trajectory (a) and its corresponding x and y zero order gradient moments (b) show that the x-gradient ($G_x$) and y-gradient ($G_y$) were balanced during acquisition. The crusher gradient ($G_z$) dephases the transverse mgnetizaton after each readout for every TR.

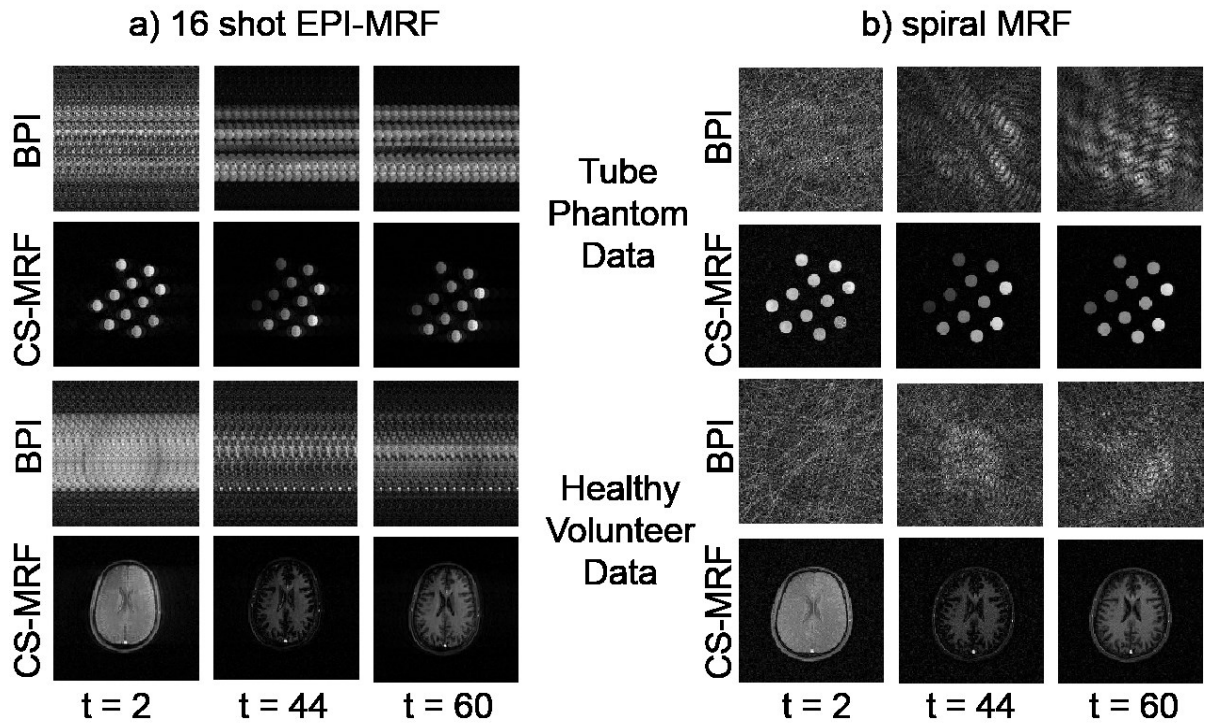

Figure 2: Figure showing the highly aliased back projected image (BPI) and Compressed sensing recovered image (CS-MRF) at different repetition indexes t for the tube phantom and healthy volunteer for a) multi-shot EPI acquisition and b) spiral acquisition

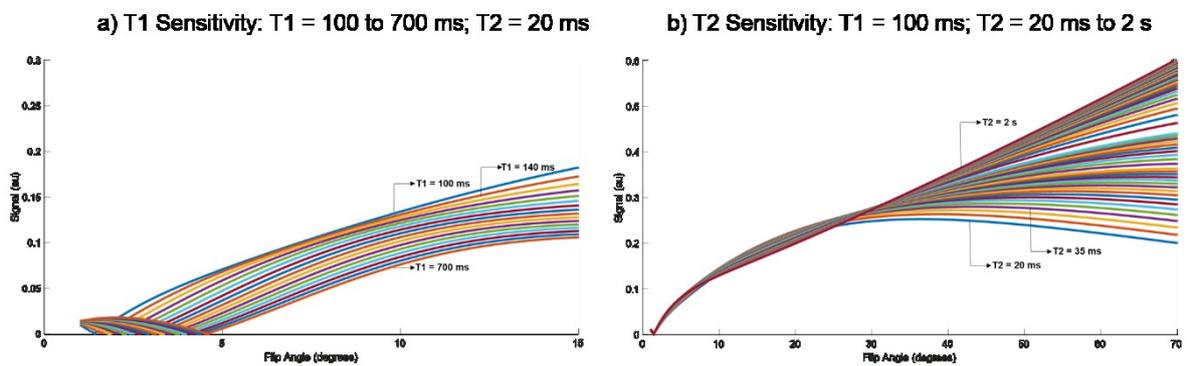

Figure 3: Figure showing the T1 and T2 sensitivity of the sequence for discriminating dictionary atoms when a linear ramp flip angle train (1° to 70°) is used during the acquisition

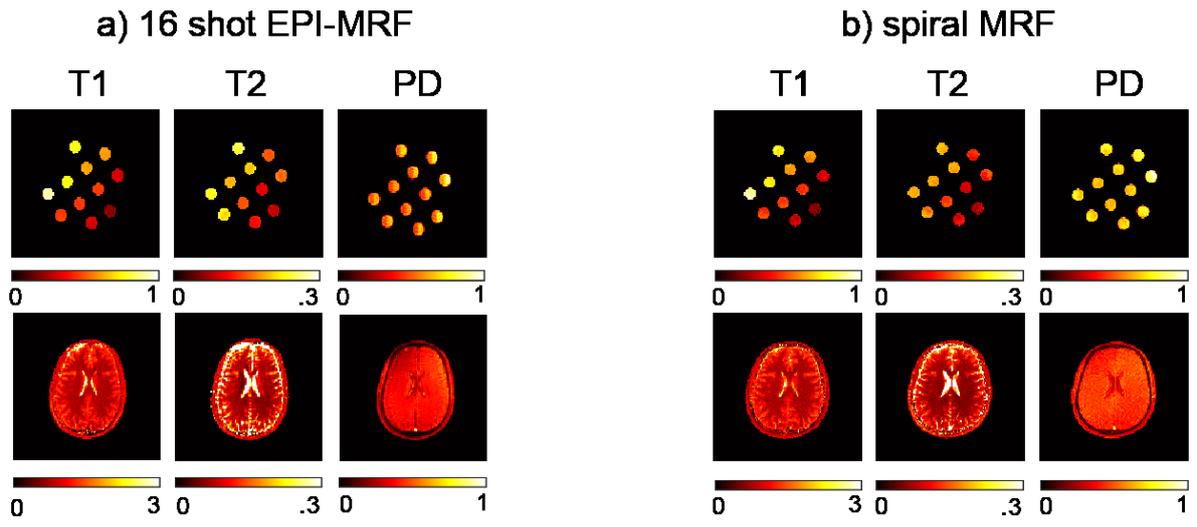

Figure 4: Figure showing the T1, T2 and PD maps for the tube phantom and healthy volunteer for a) multi-shot EPI acquisition and b) spiral acquisition

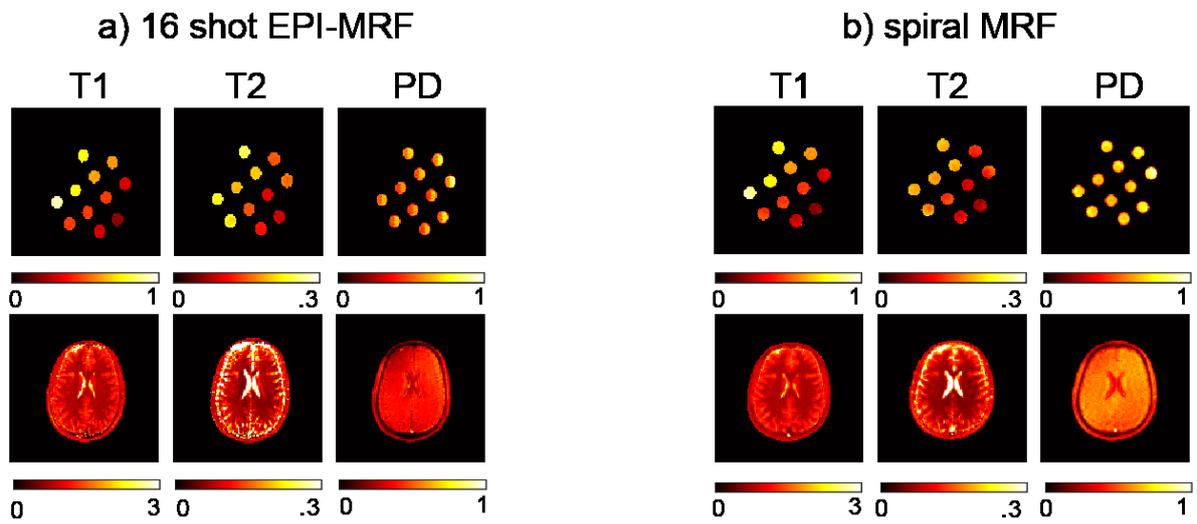

Figure 5: Figure showing the qualitatively improved T1, T2 and PD maps for the tube phantom and healthy volunteer for a) multi-shot EPI acquisition and b) spiral acquisition after the application of the BLIP algorithm. BLIP converged in fewer and cheaper iterations - 35 seconds (tube phantom) and 36 seconds (healthy volunteer) for EPI-MRF while it took 248 seconds (tube phantom) and 225 seconds (healthy volunteer) for spiral MRF.